\documentclass[11pt,a4paper,oneside]{extarticle}
\PassOptionsToPackage{unicode=true}{hyperref}
\PassOptionsToPackage{hyphens}{url}
\PassOptionsToPackage{dvipsnames,svgnames*,x11names*}{xcolor}
\usepackage{lmodern}
\usepackage{amssymb,amsmath}
\usepackage{ifpdf,ifxetex,ifluatex}
\usepackage{fixltx2e} 
\ifnum 0\ifxetex 1\fi\ifluatex 1\fi=0 
  \usepackage[T1]{fontenc}
  \usepackage[utf8]{inputenc}
  \usepackage{textcomp} 
\else 
  \usepackage{unicode-math}
  \defaultfontfeatures{Ligatures=TeX,Scale=MatchLowercase}
\fi
\IfFileExists{upquote.sty}{\usepackage{upquote}}{}
\IfFileExists{microtype.sty}{%
\usepackage[]{microtype}
\UseMicrotypeSet[protrusion]{basicmath} 
}{}
\usepackage{indentfirst}
\usepackage{xcolor}
\usepackage{hyperref}
\hypersetup{
            pdftitle={Programming with Applicative-like expressions},
            pdfauthor={Jan Malakhovski and Sergei Soloviev},
            pdfkeywords={applicative functors, algebraic data types, metaprogramming, stack machines, logic, category theory},
            breaklinks=true,
            }
\ifnum 0\ifpdf 1\fi\ifxetex 1\fi\ifluatex 1\fi=0 
\hypersetup{
            colorlinks=true,
            linkcolor=Maroon,
            citecolor=Blue,
            urlcolor=Blue,
            }
\else 
\hypersetup{
            }
\fi
\usepackage[margin=2cm]{geometry}
\usepackage{color}
\usepackage{fancyvrb}

\newcommand{\VERB}{\Verb[commandchars=\\\{\}]}
\DefineVerbatimEnvironment{Highlighting}{Verbatim}{commandchars=\\\{\}}
\newenvironment{Shaded}{}{}

\newcommand{\CommentTok}[1]{\textcolor[rgb]{0.38,0.63,0.69}{\textit{#1}}}

\newcommand{\DataTypeTok}[1]{\textcolor[rgb]{0.56,0.13,0.00}{#1}}
\newcommand{\DecValTok}[1]{\textcolor[rgb]{0.25,0.63,0.44}{#1}}

\newcommand{\FunctionTok}[1]{\textcolor[rgb]{0.02,0.16,0.49}{#1}}

\newcommand{\KeywordTok}[1]{\textcolor[rgb]{0.00,0.44,0.13}{\textbf{#1}}}
\newcommand{\NormalTok}[1]{#1}
\newcommand{\OperatorTok}[1]{\textcolor[rgb]{0.40,0.40,0.40}{#1}}
\newcommand{\OtherTok}[1]{\textcolor[rgb]{0.00,0.44,0.13}{#1}}

\newcommand{\StringTok}[1]{\textcolor[rgb]{0.25,0.44,0.63}{#1}}

\ifx\paragraph\undefined\else
\let\oldparagraph\paragraph
\renewcommand{\paragraph}[1]{\oldparagraph{#1}\mbox{}}
\fi
\ifx\subparagraph\undefined\else
\let\oldsubparagraph\subparagraph
\renewcommand{\subparagraph}[1]{\oldsubparagraph{#1}\mbox{}}
\fi

\makeatletter
\def\fps@figure{htbp}
\makeatother

\usepackage[style=numeric-comp,backend=biber,natbib=true,backref=true,sorting=none]{biblatex}
\usepackage{attachfile2}

\def\ignore#1{}

\usepackage{enumitem}

\usepackage{amsthm}

\newtheorem{remark}{Remark}

\usepackage{cleveref}
\usepackage{footnotebackref}
\crefformat{footnote}{#2\footnotemark[#1]#3}
\usepackage[affil-it]{authblk}
\author{Jan Malakhovski and Sergei Soloviev\thanks{\href{mailto:papers@oxij.org}{papers@oxij.org}; preferably with paper title in the subject line}}
\affil{IRIT, University of Toulouse-3 and ITMO University}

\title{Programming with Applicative-like expressions}

\date{2015 - 2019}

\begin{document}
\maketitle

\begin{abstract}

The fact that \VERB|\DataTypeTok{Applicative}| type class allows one to
express simple parsers in a variable-less combinatorial style is well
appreciated among Haskell programmers for its conceptual simplicity,
ease of use, and usefulness for semi-automated code generation
(metaprogramming).

We notice that such \VERB|\DataTypeTok{Applicative}| computations can be
interpreted as providing a \emph{mechanism} to construct a data type
with ``ports'' ``pluggable'' by subcomputations. We observe that it is
this property that makes them so much more convenient in practice than
the usual way of building the same computations using conventional
composition. We distill this observation into a more general algebraic
structure of (and/or technique for expressing)
``\VERB|\DataTypeTok{Applicative}|-like'' computations and demonstrate
several other instances of this structure.

Our interest in all of this comes from the fact that the aforementioned
instances allow us to express arbitrary transformations between simple
data types of a single constructor (similarly to how
\VERB|\DataTypeTok{Applicative}| parsing allows to transform from
streams of \VERB|\DataTypeTok{Char}|s to such data types) using a style
that closely follows conventional \VERB|\DataTypeTok{Applicative}|
computations, thus greatly simplifying (if not completely automating
away) a lot of boiler-plate code present in many functional programs.

\end{abstract}

\small\tableofcontents\normalsize

\hypertarget{preliminaries}{%
\section{Preliminaries}\label{preliminaries}}

This paper describes an algebraic structure (and/or a technique) for
expressing certain kinds of computations. The presented derivation of
said structure (technique) starts from observing Haskell's
\VERB|\DataTypeTok{Applicative}| type class~\cite{mcbride-paterson-08}
and then generalizing it. The result, however, is language-agnostic (the
same way \VERB|\DataTypeTok{Applicative}| is, as both structures can be
applied to most functional programming languages in some shape or form,
even if a language in question can not explicitly express required type
signatures) and theoretically interesting (as it points to several
curious connections to category theory and logic).

Since the idea grows from Haskell and most related literature uses
Haskell, it is natural to express the derivation of the structure
(technique) in Haskell. Therefore, this paper is organized as a series
of Literate Haskell programs in a single Emacs Org-Mode
tree~\cite{OrgMode, Schulte:2011:MLCELPRR} (then, most likely, compiled
into the representation you are looking at right now).\footnote{\label{fn:source}
  The source code is available at
  \url{https://oxij.org/paper/ApplicativeLike/}.
  \ifnum 0\ifpdf 1\fi\ifxetex 1\fi\ifluatex 1\fi=0 It also gets embedded
  straight into the PDF version of the article when compiled with a
  modern TeX engine. Unfortunately, the file you are looking at was
  compiled using \texttt{dvipdf}. Properly compiled version is available
  via the above link. \else
  It is also embedded straight into the PDF version of this article (click here \attachfile{article.org} or look for "attachments" in
  your PDF viewer).
  \fi All runnable code in the paper was tested with GHC~\cite{GHC}
  version 8.6.} Moreover, for uniformity reasons we shall also use
Haskell type class names for the names of the corresponding algebraic
structures where appropriate (e.g. ``\VERB|\DataTypeTok{Applicative}|''
instead of ``applicative functor'') as not to cause any confusion
between the code and the rest of the text.

\hypertarget{introduction}{%
\section{Introduction}\label{introduction}}

Let us recall the definition of \VERB|\DataTypeTok{Applicative}| type
class~\cite{mcbride-paterson-08} as it is currently defined in the
\texttt{base}~\cite{Hackage:base4900} package of Hackage~\cite{Hackage}

\begin{Shaded}
\begin{Highlighting}[]
\KeywordTok{infixl} \DecValTok{4} \OperatorTok{<*>}
\KeywordTok{class} \DataTypeTok{Functor}\NormalTok{ f }\OtherTok{=>} \DataTypeTok{Applicative}\NormalTok{ f }\KeywordTok{where}
\OtherTok{  pure ::}\NormalTok{ a }\OtherTok{->}\NormalTok{ f a}
\OtherTok{  (<*>) ::}\NormalTok{ f (a }\OtherTok{->}\NormalTok{ b) }\OtherTok{->}\NormalTok{ f a }\OtherTok{->}\NormalTok{ f b}
\end{Highlighting}
\end{Shaded}

One can think of the above definition as simply providing a generic
``constant injector'' \VERB|\FunctionTok{pure}| and a somewhat generic
``function application''
\VERB|\NormalTok{(}\OperatorTok{<*>}\NormalTok{)}| operator. (The
referenced \VERB|\DataTypeTok{Functor}| type class and any related
algebraic laws can be completely ignored for the purposes of this
article.) For instance, an identity on Haskell types is obviously an
\VERB|\DataTypeTok{Applicative}| with
\VERB|\FunctionTok{pure} \OtherTok{=} \FunctionTok{id}| and
\VERB|\NormalTok{(}\OperatorTok{<*>}\NormalTok{)}| being the
conventional function application (the one that is usually denoted by
simple juxtaposition of terms), but there are many more complex
instances of this type class
(see~\cite{HaskellWiki:Typeclassopedia, Malakhovski:2018:EME} for
comprehensive overviews of this and related algebraic structures), most
(for the purposes of this article) notably, including
\VERB|\DataTypeTok{Applicative}| parsing combinators.

Those are very popular in practice as they simplify parsing of simple
data types (``simple'' in this context means ``without any type or data
dependencies between different parts'') to the point of triviality. For
instance, given appropriate \VERB|\DataTypeTok{Applicative}| parsing
machinery like Parsec~\cite{Hackage:parsec3111},
Attoparsec~\cite{Hackage:attoparsec01310} or
Megaparsec~\cite{Hackage:megaparsec630} one can parse a simple data type
like

\begin{Shaded}
\begin{Highlighting}[]
\KeywordTok{data} \DataTypeTok{Device} \OtherTok{=} \DataTypeTok{Device}
\NormalTok{  \{}\OtherTok{ block ::} \DataTypeTok{Bool}
\NormalTok{  ,}\OtherTok{ major ::} \DataTypeTok{Int}
\NormalTok{  ,}\OtherTok{ minor ::} \DataTypeTok{Int}\NormalTok{ \}}

\OtherTok{exampleDevice ::} \DataTypeTok{Device}
\NormalTok{exampleDevice }\OtherTok{=} \DataTypeTok{Device} \DataTypeTok{False} \DecValTok{19} \DecValTok{1}
\end{Highlighting}
\end{Shaded}

\noindent from a straightforward serialized representation with just

\begin{Shaded}
\begin{Highlighting}[]
\KeywordTok{class} \DataTypeTok{Parsable}\NormalTok{ a }\KeywordTok{where}
\OtherTok{  parse ::} \DataTypeTok{Parser}\NormalTok{ a}

\KeywordTok{instance} \DataTypeTok{Parsable} \DataTypeTok{Device} \KeywordTok{where}
\NormalTok{  parse }\OtherTok{=} \FunctionTok{pure} \DataTypeTok{Device} \OperatorTok{<*>}\NormalTok{ parse }\OperatorTok{<*>}\NormalTok{ parse }\OperatorTok{<*>}\NormalTok{ parse}
\end{Highlighting}
\end{Shaded}

While clearly limited to simple data types of a single\footnote{Two or
  more constructors can be handled with the help of
  \VERB|\DataTypeTok{Alternative}| type class and some tagging of
  choices, but that is out of scope of this article.} constructor, this
approach is very useful in practice. Firstly, since these kinds of
expressions make no variable bindings and all they do is repeatedly
apply \VERB|\NormalTok{parse}| it is virtually impossible to make a
mistake. Secondly, for the same reason it is exceptionally easy to
generate such expressions via Template Haskell and similar
metaprogramming mechanisms. Which is why a plethora of Hackage libraries
use this approach.

In this paper we shall demonstrate a surprisingly simple technique that
can be used to make computations expressing arbitrary transformations
between simple data types of a single constructor while keeping the
general form of \VERB|\DataTypeTok{Applicative}| expressions as they
were shown above. Since we design our expressions to look similar to
those produced with the help of \VERB|\DataTypeTok{Applicative}| type
class but the underlying structure is not
\VERB|\DataTypeTok{Applicative}| we shall call them
``\VERB|\DataTypeTok{Applicative}|-like''. \Cref{sec:examples} provides
some motivating examples that show why we want to use
\VERB|\DataTypeTok{Applicative}|-like computations to express
transformations between data types. \Cref{sec:definition} formalizes the
notion of ``\VERB|\DataTypeTok{Applicative}|-like'' and discusses the
properties we expect from such expressions.
\Cref{sec:deriving-the-technique} derives one particular structure for
one of the motivating examples using LISP-encoding for deconstructing
data types. \Cref{sec:implementation} proceeds to derive the rest of
motivating examples by applying the same idea, thus showing that
\cref{sec:deriving-the-technique} describes a technique, not an isolated
example. \Cref{sec:implementation} ends by demonstrating the total
expressive power of the technique. \Cref{sec:scott} repeats the
derivation and the implementations for Scott-encoded data types.
\Cref{sec:general-case} observes the general structure behind all of the
terms used in the paper. \Cref{sec:formally} gives a formal description
of the technique and the underlying general algebraic structure.
\Cref{sec:conclusion} refers to related work and wraps everything up.

\hypertarget{motivating-examples}{%
\section{Motivating examples}\label{motivating-examples}}

\label{sec:examples}

Consider the following expressions produced with the help of first
author's favorite \texttt{safecopy}~\cite{Hackage:safecopy0943}
data-type-to-binary serialization-deserialization library which can be
used to deserialize-serialize \VERB|\DataTypeTok{Device}| with the
following code snippet (simplified\footnote{The actual working code for
  the actual library looks a bit more complex, but the \texttt{safecopy}
  library also provides Template Haskell functions that derive these
  \VERB|\DataTypeTok{SafeCopy}| instances automatically, so, in
  practice, one would not need to write this code by hand in any case.})

\begin{Shaded}
\begin{Highlighting}[]
\KeywordTok{instance} \DataTypeTok{SafeCopy} \DataTypeTok{Device} \KeywordTok{where}
\NormalTok{  getCopy }\OtherTok{=} \FunctionTok{pure} \DataTypeTok{Device} \OperatorTok{<*>}\NormalTok{ getCopy }\OperatorTok{<*>}\NormalTok{ getCopy }\OperatorTok{<*>}\NormalTok{ getCopy}

\NormalTok{  putCopy (}\DataTypeTok{Device}\NormalTok{ b x y) }\OtherTok{=}\NormalTok{ putCopy b }\OperatorTok{>>}\NormalTok{ putCopy x }\OperatorTok{>>}\NormalTok{ putCopy y}
\end{Highlighting}
\end{Shaded}

Note that while \VERB|\NormalTok{getCopy}| definition above is trivial,
\VERB|\NormalTok{putCopy}| definition binds variables. Would not it be
better if we had an \VERB|\DataTypeTok{Applicative}|-like machinery with
which we could rewrite \VERB|\NormalTok{putCopy}| into something like

\begin{Shaded}
\begin{Highlighting}[]
\NormalTok{putCopy }\OtherTok{=}\NormalTok{ depure unDevice }\OperatorTok{<**>}\NormalTok{ putCopy }\OperatorTok{<**>}\NormalTok{ putCopy }\OperatorTok{<**>}\NormalTok{ putCopy}
\end{Highlighting}
\end{Shaded}

\noindent which, incidentally, would also allow us to generate both
functions from a single expression? This idea does not feel like a big
stretch of imagination for several reasons:

\begin{itemize}
\item
  there are libraries that can do both parsing and pretty printing using
  a single expression, e.g.~\cite{Hackage:syntax1000},
\item
  the general pattern of \VERB|\NormalTok{putCopy}| feels very similar
  to computations in \VERB|\NormalTok{(}\OtherTok{->}\NormalTok{) a}|
  (the type of ``functions from \VERB|\NormalTok{a}|'') as it, too, is a
  kind of computation in a context with a constant value, aka
  \VERB|\DataTypeTok{Reader}|
  \VERB|\DataTypeTok{Monad}|~\cite{Hackage:transformers0520}, which is
  an instance of \VERB|\DataTypeTok{Applicative}|.\footnote{\label{fn:function-reader}We
    shall utilize this fact in the following sections.}
\end{itemize}

Another example is the data-type-to-JSON-to-strings
serialization-deserialization part of
\texttt{aeson}~\cite{Hackage:aeson1420} library which gives the
following class signatures to its deserializer and serializer from/to
JSON respectively.

\begin{Shaded}
\begin{Highlighting}[]
\KeywordTok{class} \DataTypeTok{FromJSON}\NormalTok{ a }\KeywordTok{where}
\OtherTok{  parseJSON ::} \DataTypeTok{Value} \OtherTok{->} \DataTypeTok{Parser}\NormalTok{ a}

\KeywordTok{class} \DataTypeTok{ToJSON}\NormalTok{ a }\KeywordTok{where}
\OtherTok{  toJSON ::}\NormalTok{ a }\OtherTok{->} \DataTypeTok{Value}
\end{Highlighting}
\end{Shaded}

In the above, \VERB|\DataTypeTok{Value}| is a JSON value and
\VERB|\DataTypeTok{Parser}\NormalTok{ a}| is a Scott-transformed
variation of
\VERB|\DataTypeTok{Either} \DataTypeTok{ErrorMessage}\NormalTok{ a}|.
Assuming \VERB|\NormalTok{(}\OperatorTok{.:}\NormalTok{)}| to be a
syntax sugar for \VERB|\FunctionTok{lookup}|-in-a-map-by-name function
and \VERB|\NormalTok{(}\OperatorTok{.=}\NormalTok{)}| a pair
constructor, we can give the following instances for the
\VERB|\DataTypeTok{Device}| data type by emulating examples given in the
package's own documentation

\begin{Shaded}
\begin{Highlighting}[]
\KeywordTok{instance} \DataTypeTok{FromJSON} \DataTypeTok{Device} \KeywordTok{where}
\NormalTok{  parseJSON (}\DataTypeTok{Object}\NormalTok{ v) }\OtherTok{=} \FunctionTok{pure} \DataTypeTok{Device}
                     \OperatorTok{<*>}\NormalTok{ v }\OperatorTok{.:} \StringTok{"block"}
                     \OperatorTok{<*>}\NormalTok{ v }\OperatorTok{.:} \StringTok{"major"}
                     \OperatorTok{<*>}\NormalTok{ v }\OperatorTok{.:} \StringTok{"minor"}
\NormalTok{  parseJSON _          }\OtherTok{=}\NormalTok{ empty}

\KeywordTok{instance} \DataTypeTok{ToJSON} \DataTypeTok{Device} \KeywordTok{where}
\NormalTok{  toJSON (}\DataTypeTok{Device}\NormalTok{ b x y) }\OtherTok{=}\NormalTok{ object}
\NormalTok{                      [ }\StringTok{"block"} \OperatorTok{.=}\NormalTok{ b}
\NormalTok{                      , }\StringTok{"major"} \OperatorTok{.=}\NormalTok{ x}
\NormalTok{                      , }\StringTok{"minor"} \OperatorTok{.=}\NormalTok{ y ]}
\end{Highlighting}
\end{Shaded}

Note that here, again, we have to bind variables in
\VERB|\NormalTok{toJSON}|. Moreover, note that in this example even
\VERB|\NormalTok{parseJSON}| underuses the
\VERB|\DataTypeTok{Applicative}| structure by ignoring the fact that
\VERB|\DataTypeTok{Value}| can be packed into
\VERB|\DataTypeTok{Parser}| by making the latter into a
\VERB|\DataTypeTok{Reader}|.\footnote{As noted under
  footnote~\ref{fn:function-reader} and demonstrated in detail in
  \cref{sec:deriving-the-technique}. However, this underuse has a
  reasonable explanation for \texttt{aeson}:
  \VERB|\DataTypeTok{Value}|'s definition is \emph{too structured} to
  have a conventional parser combinator library that can make this trick
  work in the general case (i.e. not just in the above example). This
  problem can be solved using indexed \VERB|\DataTypeTok{Monad}|ic
  parser combinators but that is out of scope of this article.}

Other serialization-deserialization problems, e.g. conventional
pretty-printing with the standard \VERB|\DataTypeTok{Show}| type
class~\cite{Hackage:base4900} are, of course, the instances of the same
pattern, as we shall demonstrate in the following sections.

Finally, as a bit more involved example, imagine an application that
benchmarks some other software applications on given inputs, records
logs they produce and then computes per-application averages

\begin{Shaded}
\begin{Highlighting}[]
\KeywordTok{data} \DataTypeTok{Benchmark}\NormalTok{ a }\OtherTok{=} \DataTypeTok{Benchmark}
\NormalTok{  \{}\OtherTok{ firstApp ::}\NormalTok{ a}
\NormalTok{  ,}\OtherTok{ firstLog ::} \DataTypeTok{String}
\NormalTok{  ,}\OtherTok{ secondApp ::}\NormalTok{ a}
\NormalTok{  ,}\OtherTok{ secondLog ::} \DataTypeTok{String}
\NormalTok{  \}}

\KeywordTok{type} \DataTypeTok{Argv}    \OtherTok{=}\NormalTok{ [}\DataTypeTok{String}\NormalTok{]}
\KeywordTok{type} \DataTypeTok{Inputs}  \OtherTok{=} \DataTypeTok{Benchmark} \DataTypeTok{Argv}
\KeywordTok{type} \DataTypeTok{Outputs} \OtherTok{=} \DataTypeTok{Benchmark} \DataTypeTok{Integer}
\KeywordTok{type} \DataTypeTok{Avgs}    \OtherTok{=} \DataTypeTok{Benchmark} \DataTypeTok{Double}
\end{Highlighting}
\end{Shaded}

\begin{Shaded}
\begin{Highlighting}[]
\OtherTok{benchmark ::} \DataTypeTok{Inputs} \OtherTok{->} \DataTypeTok{IO} \DataTypeTok{Outputs}
\OtherTok{average ::}\NormalTok{ [ }\DataTypeTok{Outputs}\NormalTok{ ] }\OtherTok{->} \DataTypeTok{Avgs}
\end{Highlighting}
\end{Shaded}

Assuming that we have aforementioned machinery for
\VERB|\DataTypeTok{SafeCopy}| we can trivially autogenerate all of the
needed glue code to deserialize \VERB|\DataTypeTok{Inputs}|, serialize
\VERB|\DataTypeTok{Outputs}| and \VERB|\DataTypeTok{Avgs}|. The
\VERB|\NormalTok{benchmark}| is the core of our application, so let us
assume that it is not trivial to autogenerate and we have to write it by
hand. We are now left with the ``\VERB|\NormalTok{average}|'' function.
Let us assume that for the numeric parts of the
\VERB|\DataTypeTok{Outputs}| type it is just a \VERB|\NormalTok{fold}|
with point-wise sum over the list of \VERB|\DataTypeTok{Outputs}|
followed by a point-wise divide by their \VERB|\FunctionTok{length}| and
for the \VERB|\DataTypeTok{String}| parts it simply point-wise
concatenates all the logs.

Now, do we really want to write those binary operators completely by
hand? Note that this \VERB|\DataTypeTok{Benchmark}| example was
carefully crafted: it is not self- or mutually-recursive and, at the
same time, it is also not particularly homogeneous as different fields
require different operations. In other words, things like
SYB~\cite{Laemmel:2003:SYB}, Uniplate~\cite{Mitchell:2007:Uniplate},
Multiplate~\cite{Hackage:multiplate003} or
Lenses~\cite{Kmett:Lens, Hackage:lens417} are not particularly useful in
this case.\footnote{Strictly speaking, both operations used in the
  ``sum'' part of ``\VERB|\NormalTok{average}|'' are
  \VERB|\DataTypeTok{Monoid}| operators, so generalized
  \VERB|\FunctionTok{zip}|s provided by some of the mentioned libraries
  can be used to implement that part, but the ``divide'' part is not so
  homogeneous.} Of course, in this particular example, it is possible to
distill the computation pattern into something like

\begin{Shaded}
\begin{Highlighting}[]
\OtherTok{lift2B ::}\NormalTok{ (a }\OtherTok{->}\NormalTok{ b }\OtherTok{->}\NormalTok{ c) }\OtherTok{->}\NormalTok{ (}\DataTypeTok{Benchmark}\NormalTok{ a }\OtherTok{->} \DataTypeTok{Benchmark}\NormalTok{ b }\OtherTok{->} \DataTypeTok{Benchmark}\NormalTok{ c)}
\NormalTok{lift2B f (}\DataTypeTok{Benchmark}\NormalTok{ a1 l1 a2 l2) (}\DataTypeTok{Benchmark}\NormalTok{ b1 l3 b2 l4)}
  \OtherTok{=} \DataTypeTok{Benchmark}\NormalTok{ (f a1 b1) (l1 }\OperatorTok{++}\NormalTok{ l3) (f a2 b2) (l2 }\OperatorTok{++}\NormalTok{ l4)}
\end{Highlighting}
\end{Shaded}

\noindent and then use \VERB|\NormalTok{lift2B}| to implement both
functions (with some unsightly hackery for the division part), but would
not it be even better if instead we had an
\VERB|\DataTypeTok{Applicative}|-like machinery that would allow us to
write the \VERB|\NormalTok{average}| function directly, such as

\begin{Shaded}
\begin{Highlighting}[]
\NormalTok{average ls }\OtherTok{=}\NormalTok{ runMap }\OperatorTok{$}\NormalTok{ bdivide folded }\KeywordTok{where}
\NormalTok{  len }\OtherTok{=} \FunctionTok{fromIntegral} \OperatorTok{$} \FunctionTok{length}\NormalTok{ ls}
\NormalTok{  avg }\OtherTok{=}\NormalTok{ ((}\OperatorTok{/}\NormalTok{ len) }\OperatorTok{.} \FunctionTok{fromIntegral}\NormalTok{)}

\NormalTok{  bappend }\OtherTok{=}\NormalTok{ depureZip }\DataTypeTok{Benchmark}\NormalTok{ unBenchmark unBenchmark}
    \OtherTok{`zipa`}\NormalTok{ (}\OperatorTok{+}\NormalTok{) }\OtherTok{`zipa`}\NormalTok{ (}\OperatorTok{++}\NormalTok{)}
    \OtherTok{`zipa`}\NormalTok{ (}\OperatorTok{+}\NormalTok{) }\OtherTok{`zipa`}\NormalTok{ (}\OperatorTok{++}\NormalTok{)}

\NormalTok{  folded }\OtherTok{=}\NormalTok{ foldl' (\textbackslash{}a b }\OtherTok{->}\NormalTok{ runZip }\OperatorTok{$}\NormalTok{ bappend a b)}
\NormalTok{                  (}\DataTypeTok{Benchmark} \DecValTok{0} \StringTok{""} \DecValTok{0} \StringTok{""}\NormalTok{) ls}

\NormalTok{  bdivide }\OtherTok{=}\NormalTok{ depureMap }\DataTypeTok{Benchmark}\NormalTok{ unBenchmark}
    \OtherTok{`mapa`}\NormalTok{ avg }\OtherTok{`mapa`} \FunctionTok{id}
    \OtherTok{`mapa`}\NormalTok{ avg }\OtherTok{`mapa`} \FunctionTok{id}
\end{Highlighting}
\end{Shaded}

\noindent similarly to how we would solve similar problems over
homogeneous lists?

\hypertarget{problem-definition}{%
\section{Problem definition}\label{problem-definition}}

\label{sec:definition}

Before going into derivation of the actual implementation let us
describe what we mean by ``\VERB|\DataTypeTok{Applicative}|-like'' more
precisely.

Note that the type of \VERB|\NormalTok{(}\OperatorTok{<*>}\NormalTok{)}|
operator of \VERB|\DataTypeTok{Applicative}|

\begin{Shaded}
\begin{Highlighting}[]
\OtherTok{(<*>) ::}\NormalTok{ f (a }\OtherTok{->}\NormalTok{ b) }\OtherTok{->}\NormalTok{ f a }\OtherTok{->}\NormalTok{ f b}
\end{Highlighting}
\end{Shaded}

\noindent at least in the context of constructing data types (of which
\VERB|\DataTypeTok{Applicative}| parsers are a prime example), can be
generalized and reinterpreted as

\begin{Shaded}
\begin{Highlighting}[]
\OtherTok{plug ::}\NormalTok{ f full }\OtherTok{->}\NormalTok{ g piece }\OtherTok{->}\NormalTok{ f fullWithoutThePiece}
\end{Highlighting}
\end{Shaded}

\noindent where

\begin{itemize}
\item
  \VERB|\NormalTok{f full}| is a computation that \emph{provides a
  mechanism} to handle the \VERB|\NormalTok{full}| structure,
\item
  \VERB|\NormalTok{g piece}| is another kind of computation that
  \emph{actually handles} a \VERB|\NormalTok{piece}| of the
  \VERB|\NormalTok{full}| structure
  (\VERB|\NormalTok{g }\OperatorTok{==}\NormalTok{ f}| for
  \VERB|\DataTypeTok{Applicative}| parsers, of course),
\item
  and \VERB|\NormalTok{f fullWithoutThePiece}| is a computation that
  provided a mechanism to handle the leftover part.
\end{itemize}

Note that this interpretation, in some sense, reverses conventional
wisdom on how such transformations are usually expressed.

For instance, conventionally, to parse (pretty-print, etc) some
structure one first makes up computations that handle
\VERB|\NormalTok{piece}|s and then composes them into a computation that
handles the \VERB|\NormalTok{full}| structure, i.e.

\begin{Shaded}
\begin{Highlighting}[]
\OtherTok{compose  ::}\NormalTok{ f fullWithoutThePiece }\OtherTok{->}\NormalTok{ g piece }\OtherTok{->}\NormalTok{ f full}
\CommentTok{-- or}
\OtherTok{compose' ::}\NormalTok{ g piece }\OtherTok{->}\NormalTok{ f fullWithoutThePiece }\OtherTok{->}\NormalTok{ f full}
\end{Highlighting}
\end{Shaded}

Meanwhile, \VERB|\DataTypeTok{Applicative}|-like expressions, in some
sense, work backwards: they provide up a mechanism to handle (parse,
pretty-print, etc) the \VERB|\NormalTok{full}| structure that exposes
``ports'' that subcomputations \VERB|\NormalTok{plug}| with computations
that handle different \VERB|\NormalTok{piece}|s.

\begin{remark}

It is rather interesting to think about the conventional function
application in these terms: it describes a way to make a computation
that produces \VERB|\NormalTok{b}| given a mechanism to construct a
partial version of \VERB|\NormalTok{b}| denoted as
\VERB|\NormalTok{a }\OtherTok{->}\NormalTok{ b}| by plugging its only
port with a computation that produces \VERB|\NormalTok{a}|. In other
words, this outlook is a reminder that functions can be seen as goals,
the same way Haskell's type class instance inference (or Prolog) does.
Moreover, note that while such a description sounds obvious for a lazy
language, it is also a reminder that, in general, there is a distinction
between values and computations.

\end{remark}

To summarize, the crucial part of \VERB|\DataTypeTok{Applicative}|-like
computations is the fact that they compose subcomputations in reverse
order w.r.t. the types they handle. This reversal is the cornerstone
that provides three important properties:

\begin{itemize}
\item
  A sequence of subcomputations in an expression matches the sequence of
  parts in the corresponding data type.
\item
  A top-level computation can decide on all data types \emph{first} and
  then delegate handing of parts to subcomputations without worrying
  about reassembling their results (which is why we say it ``provides a
  mechanism'' that subcomputations use).
\item
  As a consequence, in the presence of type inference, a mechanism for
  ad-hoc polymorphism (be it type classes, like in Haskell, or something
  else) can be used to automatically select implementations matching
  corresponding \VERB|\NormalTok{piece}|s.
\end{itemize}

It is the combination of these three properties that makes
\VERB|\DataTypeTok{Applicative}|-like expressions (including
\VERB|\DataTypeTok{Applicative}| parsers) so convenient in practice.

\hypertarget{deriving-the-technique}{%
\section{Deriving the technique}\label{deriving-the-technique}}

\label{sec:deriving-the-technique}

We shall now demonstrate the derivation of the main technique of the
paper. Before we start, let us encode reverses to
\VERB|\DataTypeTok{Device}| and \VERB|\DataTypeTok{Benchmark}|
constructors (i.e. ``destructors'') using the LISP-encoding (see below
for motivation, an alternative approach using Scott-encoding is
discussed in section~\ref{sec:scott}).

\begin{Shaded}
\begin{Highlighting}[]
\OtherTok{unDeviceLISP ::} \DataTypeTok{Device} \OtherTok{->}\NormalTok{ (}\DataTypeTok{Bool}\NormalTok{, (}\DataTypeTok{Int}\NormalTok{, (}\DataTypeTok{Int}\NormalTok{, ())))}
\NormalTok{unDeviceLISP (}\DataTypeTok{Device}\NormalTok{ b x y) }\OtherTok{=}\NormalTok{ (b, (x, (y, ())))}

\OtherTok{unBenchmarkLISP ::} \DataTypeTok{Benchmark}\NormalTok{ a }\OtherTok{->}\NormalTok{ (a, (}\DataTypeTok{String}\NormalTok{, (a, (}\DataTypeTok{String}\NormalTok{, ()))))}
\NormalTok{unBenchmarkLISP (}\DataTypeTok{Benchmark}\NormalTok{ a b c d) }\OtherTok{=}\NormalTok{ (a, (b, (c, (d, ()))))}
\end{Highlighting}
\end{Shaded}

Now, let us start by deriving an \VERB|\DataTypeTok{Applicative}|-like
pretty-printer for \VERB|\DataTypeTok{Device}|. The target expression is
as follows

\begin{Shaded}
\begin{Highlighting}[]
\NormalTok{showDevice }\OtherTok{=}\NormalTok{ depureShow unDeviceLISP }\OtherTok{`showa`} \FunctionTok{show}
                                     \OtherTok{`showa`} \FunctionTok{show}
                                     \OtherTok{`showa`} \FunctionTok{show}
\end{Highlighting}
\end{Shaded}

Remember that the type pattern for the \VERB|\NormalTok{plug}| operator
from the previous section

\begin{Shaded}
\begin{Highlighting}[]
\OtherTok{plug ::}\NormalTok{ f full }\OtherTok{->}\NormalTok{ g piece }\OtherTok{->}\NormalTok{ f fullWithoutThePiece}
\end{Highlighting}
\end{Shaded}

\noindent already prescribes a certain way of implementing the missing
operators. Firstly, if we follow the logic for parsing, the
\VERB|\NormalTok{f}| type-level function should construct a type that
contains some internal state. Secondly, the rest of the expression
clearly requires \VERB|\NormalTok{depureShow}| to generate the initial
state and \VERB|\NormalTok{showa}| to transform the internal state while
chopping away at the parts of the \VERB|\DataTypeTok{Device}|.

Let us simplify the task of deriving these functions by writing out the
desired type and making \VERB|\DataTypeTok{Device}| argument explicit.
Let us also apply the result of the whole computation to
\VERB|\NormalTok{runShow}| function to lift the restriction on the
return type.

\begin{Shaded}
\begin{Highlighting}[]
\OtherTok{showDevice' ::} \DataTypeTok{Device} \OtherTok{->} \DataTypeTok{String}
\NormalTok{showDevice' d }\OtherTok{=}\NormalTok{ runShow }\OperatorTok{$}\NormalTok{ depureShow' (unDeviceLISP d) }\OtherTok{`showa'`} \FunctionTok{show}
                                                       \OtherTok{`showa'`} \FunctionTok{show}
                                                       \OtherTok{`showa'`} \FunctionTok{show}
\end{Highlighting}
\end{Shaded}

What should be the type of \VERB|\NormalTok{showa'}|? Clearly, something
like

\begin{Shaded}
\begin{Highlighting}[]
\OtherTok{showa' ::}\NormalTok{ (s, (a, b)) }\OtherTok{->}\NormalTok{ (a }\OtherTok{->} \DataTypeTok{String}\NormalTok{) }\OtherTok{->}\NormalTok{ (s, b)}
\end{Highlighting}
\end{Shaded}

\noindent should work and match the type pattern of
\VERB|\NormalTok{plug}|. The
\VERB|\NormalTok{a }\OtherTok{->} \DataTypeTok{String}| part follows
from the expression itself, the \VERB|\NormalTok{(_ , (a, b))}| and
\VERB|\NormalTok{(_ , b)}| parts come from chopping away at LISP-encoded
deconstructed data type, and \VERB|\NormalTok{s}| plays the role of the
internal pretty-printing state. We just need to decide on the value of
\VERB|\NormalTok{s}|. The most simple option seems to be to the list of
\VERB|\DataTypeTok{String}|s that is to be concatenated in
\VERB|\NormalTok{runShow}|. The rest of the code pretty much writes
itself:

\begin{Shaded}
\begin{Highlighting}[]
\OtherTok{depureShow' ::}\NormalTok{ a }\OtherTok{->}\NormalTok{ ([}\DataTypeTok{String}\NormalTok{], a)}
\NormalTok{depureShow' a }\OtherTok{=}\NormalTok{ ([], a)}

\OtherTok{showa' ::}\NormalTok{ ([}\DataTypeTok{String}\NormalTok{], (a, b)) }\OtherTok{->}\NormalTok{ (a }\OtherTok{->} \DataTypeTok{String}\NormalTok{) }\OtherTok{->}\NormalTok{ ([}\DataTypeTok{String}\NormalTok{], b)}
\NormalTok{showa' (s, (a, b)) f }\OtherTok{=}\NormalTok{ ((f a)}\OperatorTok{:}\NormalTok{s, b)}

\OtherTok{runShow ::}\NormalTok{ ([}\DataTypeTok{String}\NormalTok{], b) }\OtherTok{->} \DataTypeTok{String}
\NormalTok{runShow }\OtherTok{=} \FunctionTok{concat} \OperatorTok{.}\NormalTok{ intersperse }\StringTok{" "} \OperatorTok{.} \FunctionTok{reverse} \OperatorTok{.} \FunctionTok{fst}

\OtherTok{testShowDevice' ::} \DataTypeTok{String}
\NormalTok{testShowDevice' }\OtherTok{=}\NormalTok{ showDevice' exampleDevice}
\end{Highlighting}
\end{Shaded}

Now, note that \VERB|\NormalTok{showa'}| is actually a particular case
of the more generic operator

\begin{Shaded}
\begin{Highlighting}[]
\OtherTok{chop ::}\NormalTok{ (s, (a, b)) }\OtherTok{->}\NormalTok{ (s }\OtherTok{->}\NormalTok{ a }\OtherTok{->}\NormalTok{ t) }\OtherTok{->}\NormalTok{ (t, b)}
\NormalTok{chop (s, (a, b)) f }\OtherTok{=}\NormalTok{ (f s a, b)}

\NormalTok{showa'' s f }\OtherTok{=}\NormalTok{ chop s (\textbackslash{}s a }\OtherTok{->}\NormalTok{ (f a)}\OperatorTok{:}\NormalTok{s) }\CommentTok{-- == showa'}
\end{Highlighting}
\end{Shaded}

Moreover, \VERB|\NormalTok{f}| parts of that operator can be wrapped
into the \VERB|\NormalTok{(}\OtherTok{->}\NormalTok{) r}|
\VERB|\DataTypeTok{Reader}| (see under
footnote~\ref{fn:function-reader})

\begin{Shaded}
\begin{Highlighting}[]
\OtherTok{chopR ::}\NormalTok{ (r }\OtherTok{->}\NormalTok{ (s, (a, b))) }\OtherTok{->}\NormalTok{ (s }\OtherTok{->}\NormalTok{ a }\OtherTok{->}\NormalTok{ t) }\OtherTok{->}\NormalTok{ (r }\OtherTok{->}\NormalTok{ (t, b))}
\NormalTok{chopR o f r }\OtherTok{=}\NormalTok{ chop (o r) f}
\end{Highlighting}
\end{Shaded}

\noindent thus allowing us to complete the original
\VERB|\NormalTok{showDevice}|

\begin{Shaded}
\begin{Highlighting}[]
\OtherTok{showDevice ::} \DataTypeTok{Device} \OtherTok{->}\NormalTok{ ([}\DataTypeTok{String}\NormalTok{], ())}

\OtherTok{depureShow ::}\NormalTok{ (t }\OtherTok{->}\NormalTok{ b) }\OtherTok{->}\NormalTok{ t }\OtherTok{->}\NormalTok{ ([}\DataTypeTok{String}\NormalTok{], b)}
\NormalTok{depureShow f r }\OtherTok{=}\NormalTok{ ([], f r)}

\OtherTok{showa ::}\NormalTok{ (r }\OtherTok{->}\NormalTok{ ([}\DataTypeTok{String}\NormalTok{], (a, b)))}
      \OtherTok{->}\NormalTok{ (a }\OtherTok{->} \DataTypeTok{String}\NormalTok{)}
      \OtherTok{->}\NormalTok{ (r }\OtherTok{->}\NormalTok{ ([}\DataTypeTok{String}\NormalTok{], b))}
\NormalTok{showa st f }\OtherTok{=}\NormalTok{ chopR st (\textbackslash{}s a }\OtherTok{->}\NormalTok{ (f a)}\OperatorTok{:}\NormalTok{s)}

\OtherTok{testShowDevice ::} \DataTypeTok{String}
\NormalTok{testShowDevice }\OtherTok{=}\NormalTok{ runShow }\OperatorTok{$}\NormalTok{ showDevice exampleDevice}
\end{Highlighting}
\end{Shaded}

Note that the use of the LISP-encoding (i.e. the \VERB|\NormalTok{()}|
in the tails of the deconstructed types and, hence, the use of
\VERB|\FunctionTok{fst}| in \VERB|\NormalTok{runShow}|) as opposed to
using simple stacked tuples is needed to prevent special case handling
for the last argument.

Also note that the type of the second argument to
\VERB|\NormalTok{chopR}| in the definition of \VERB|\NormalTok{showa}|
is
\VERB|\NormalTok{[}\DataTypeTok{String}\NormalTok{] }\OtherTok{->}\NormalTok{ a }\OtherTok{->}\NormalTok{ [}\DataTypeTok{String}\NormalTok{]}|
which is \VERB|\DataTypeTok{CoState}| on a list of
\VERB|\DataTypeTok{String}|s. This makes a lot of sense categorically
since \VERB|\DataTypeTok{Parser}| is a kind of
\VERB|\DataTypeTok{State}| and parsing and pretty-printing are dual.
Moreover, even the fact that \VERB|\DataTypeTok{String}| is wrapped into
a list makes sense if one is to note that the above pretty-printer
produces \emph{lexemes} instead of directly producing the output string.

The above transformation from \VERB|\NormalTok{chop}| to
\VERB|\NormalTok{chopR}| will be a common theme in the following
sections, so let us distill it into a separate operator with a very
self-descriptive type

\begin{Shaded}
\begin{Highlighting}[]
\OtherTok{homWrap ::}\NormalTok{ (s }\OtherTok{->}\NormalTok{ a }\OtherTok{->}\NormalTok{ t)}
        \OtherTok{->}\NormalTok{ (r }\OtherTok{->}\NormalTok{ s) }\OtherTok{->}\NormalTok{ a }\OtherTok{->}\NormalTok{ (r }\OtherTok{->}\NormalTok{ t)}
\NormalTok{homWrap chopper o f r }\OtherTok{=}\NormalTok{ chopper (o r) f}

\NormalTok{showa''' }\OtherTok{=}\NormalTok{ homWrap }\OperatorTok{$}\NormalTok{ \textbackslash{}st f }\OtherTok{->}\NormalTok{ chop st }\OperatorTok{$}\NormalTok{ \textbackslash{}s a }\OtherTok{->}\NormalTok{ (f a)}\OperatorTok{:}\NormalTok{s }\CommentTok{-- == showa}
\end{Highlighting}
\end{Shaded}

\hypertarget{applying-the-technique}{%
\section{Applying the technique}\label{applying-the-technique}}

\label{sec:implementation}

Turning attention back to \VERB|\NormalTok{chop}| operator, note that
both types in the state tuple can be arbitrary. For instance,
\VERB|\NormalTok{s}| can be a curried data type constructor, which
immediately allows to express an \VERB|\DataTypeTok{Applicative}|-like
step-by-step equivalent of \VERB|\FunctionTok{map}|.

\begin{Shaded}
\begin{Highlighting}[]
\OtherTok{mapa ::}\NormalTok{ (r }\OtherTok{->}\NormalTok{ (x }\OtherTok{->}\NormalTok{ y, (a, b)))}
     \OtherTok{->}\NormalTok{ (a }\OtherTok{->}\NormalTok{ x)}
     \OtherTok{->}\NormalTok{ (r }\OtherTok{->}\NormalTok{ (y, b))}
\NormalTok{mapa }\OtherTok{=}\NormalTok{ homWrap }\OperatorTok{$}\NormalTok{ \textbackslash{}st f }\OtherTok{->}\NormalTok{ chop st }\OperatorTok{$}\NormalTok{ \textbackslash{}s a }\OtherTok{->}\NormalTok{ s (f a)}

\OtherTok{depureMap ::}\NormalTok{ a }\OtherTok{->}\NormalTok{ (t }\OtherTok{->}\NormalTok{ b) }\OtherTok{->}\NormalTok{ t }\OtherTok{->}\NormalTok{ (a, b)}
\NormalTok{depureMap c f r }\OtherTok{=}\NormalTok{ (c, f r)}

\NormalTok{runMap }\OtherTok{=} \FunctionTok{fst}

\OtherTok{mapDevice ::} \DataTypeTok{Device} \OtherTok{->}\NormalTok{ (}\DataTypeTok{Device}\NormalTok{, ())}
\NormalTok{mapDevice }\OtherTok{=}\NormalTok{ depureMap }\DataTypeTok{Device}\NormalTok{ unDeviceLISP}
  \OtherTok{`mapa`} \FunctionTok{not}
  \OtherTok{`mapa`}\NormalTok{ (}\OperatorTok{+} \DecValTok{100}\NormalTok{)}
  \OtherTok{`mapa`}\NormalTok{ (}\OperatorTok{+} \DecValTok{200}\NormalTok{)}

\OtherTok{testMapDevice ::} \DataTypeTok{Device}
\NormalTok{testMapDevice }\OtherTok{=}\NormalTok{ runMap }\OperatorTok{$}\NormalTok{ mapDevice exampleDevice}
\end{Highlighting}
\end{Shaded}

Moreover, by extending \VERB|\NormalTok{chop}| with two LISP-encoded
representations and repeating the whole derivation we can express an
equivalent of \VERB|\FunctionTok{zip}|.

\begin{Shaded}
\begin{Highlighting}[]
\OtherTok{chop2 ::}\NormalTok{ (s, (a, b), (c, d))}
      \OtherTok{->}\NormalTok{ (s }\OtherTok{->}\NormalTok{ a }\OtherTok{->}\NormalTok{ c }\OtherTok{->}\NormalTok{ t)}
      \OtherTok{->}\NormalTok{ (t, b, d)}
\NormalTok{chop2 (s, (a, b), (c, d)) f }\OtherTok{=}\NormalTok{ (f s a c, b, d)}

\NormalTok{homWrap2 chopper o f ra rb }\OtherTok{=}\NormalTok{ chopper (o ra rb) f}

\OtherTok{zipa ::}\NormalTok{ (ra }\OtherTok{->}\NormalTok{ rb }\OtherTok{->}\NormalTok{ (x }\OtherTok{->}\NormalTok{ y, (a, b), (c, d)))}
     \OtherTok{->}\NormalTok{ (a }\OtherTok{->}\NormalTok{ c }\OtherTok{->}\NormalTok{ x)}
     \OtherTok{->}\NormalTok{ (ra }\OtherTok{->}\NormalTok{ rb }\OtherTok{->}\NormalTok{ (y, b, d))}
\NormalTok{zipa }\OtherTok{=}\NormalTok{ homWrap2 }\OperatorTok{$}\NormalTok{ \textbackslash{}st f }\OtherTok{->}\NormalTok{ chop2 st }\OperatorTok{$}\NormalTok{ \textbackslash{}s a b }\OtherTok{->}\NormalTok{ s (f a b)}

\OtherTok{depureZip ::}\NormalTok{ a }\OtherTok{->}\NormalTok{ (ra }\OtherTok{->}\NormalTok{ b) }\OtherTok{->}\NormalTok{ (rb }\OtherTok{->}\NormalTok{ c)}
          \OtherTok{->}\NormalTok{ ra }\OtherTok{->}\NormalTok{ rb}
          \OtherTok{->}\NormalTok{ (a, b, c)}
\NormalTok{depureZip c f g ra rb }\OtherTok{=}\NormalTok{ (c, f ra, g rb)}

\OtherTok{runZip ::}\NormalTok{ (s, a, b) }\OtherTok{->}\NormalTok{ s}
\NormalTok{runZip (s, _, _) }\OtherTok{=}\NormalTok{ s}

\OtherTok{zipDevice ::} \DataTypeTok{Device} \OtherTok{->} \DataTypeTok{Device} \OtherTok{->}\NormalTok{ (}\DataTypeTok{Device}\NormalTok{, (), ())}
\NormalTok{zipDevice }\OtherTok{=}\NormalTok{ depureZip }\DataTypeTok{Device}\NormalTok{ unDeviceLISP unDeviceLISP}
  \OtherTok{`zipa`}\NormalTok{ (}\OperatorTok{&&}\NormalTok{)}
  \OtherTok{`zipa`}\NormalTok{ (}\OperatorTok{+}\NormalTok{)}
  \OtherTok{`zipa`}\NormalTok{ (}\OperatorTok{+}\NormalTok{)}

\OtherTok{testZipDevice ::} \DataTypeTok{Device}
\NormalTok{testZipDevice }\OtherTok{=}\NormalTok{ runZip }\OperatorTok{$}\NormalTok{ zipDevice exampleDevice testMapDevice}
\end{Highlighting}
\end{Shaded}

The above transformations combined with

\begin{Shaded}
\begin{Highlighting}[]
\NormalTok{unDevice }\OtherTok{=}\NormalTok{ unDeviceLISP}
\NormalTok{unBenchmark }\OtherTok{=}\NormalTok{ unBenchmarkLISP}
\end{Highlighting}
\end{Shaded}

\noindent implement all the examples from \cref{sec:examples}, thus
solving the problem as it was originally described.

Note, however, that the above technique can be trivially extended to
\VERB|\NormalTok{chop}|ping any number of data types at the same time
and, moreover, that it is not actually required to match types or even
the numbers of arguments of different constructors and destructors used
by the desired transformations. For instance, it is trivial to implement
the usual stack machine operators, e.g.

\begin{Shaded}
\begin{Highlighting}[]
\OtherTok{homWrap0 ::}\NormalTok{ (s }\OtherTok{->}\NormalTok{ t)}
         \OtherTok{->}\NormalTok{ (r }\OtherTok{->}\NormalTok{ s) }\OtherTok{->}\NormalTok{ (r }\OtherTok{->}\NormalTok{ t)}
\NormalTok{homWrap0 chopper o r }\OtherTok{=}\NormalTok{ chopper (o r)}

\CommentTok{-- syntax sugar}
\NormalTok{andThen x f }\OtherTok{=}\NormalTok{ f x}

\OtherTok{pop ::}\NormalTok{ (r }\OtherTok{->}\NormalTok{ (s, (a, b)))}
    \OtherTok{->}\NormalTok{ (r }\OtherTok{->}\NormalTok{ (s, b))}
\NormalTok{pop }\OtherTok{=}\NormalTok{ homWrap0 }\OperatorTok{$}\NormalTok{ \textbackslash{}(s, (_, b)) }\OtherTok{->}\NormalTok{ (s, b)}

\NormalTok{push }\OtherTok{=}\NormalTok{ homWrap }\OperatorTok{$}\NormalTok{ \textbackslash{}(s, b) a }\OtherTok{->}\NormalTok{ (s, (a, b))}

\NormalTok{dup }\OtherTok{=}\NormalTok{ homWrap0 }\OperatorTok{$}\NormalTok{ \textbackslash{}(s, (a, b)) }\OtherTok{->}\NormalTok{ (s, (a, (a, b)))}
\end{Highlighting}
\end{Shaded}

\noindent and use them to express some mapping function between data
types as if Haskell was a stack machine language

\begin{Shaded}
\begin{Highlighting}[]
\OtherTok{remapDevice ::} \DataTypeTok{Device} \OtherTok{->}\NormalTok{ (}\DataTypeTok{Device}\NormalTok{, ())}
\NormalTok{remapDevice }\OtherTok{=}\NormalTok{ depureMap }\DataTypeTok{Device}\NormalTok{ unDeviceLISP}
  \OtherTok{`andThen`}\NormalTok{ pop}
  \OtherTok{`push`} \DataTypeTok{True}
  \OtherTok{`mapa`} \FunctionTok{id}
  \OtherTok{`andThen`}\NormalTok{ pop}
  \OtherTok{`andThen`}\NormalTok{ dup}
  \OtherTok{`mapa`} \FunctionTok{id}
  \OtherTok{`mapa`} \FunctionTok{id}

\OtherTok{testRemapDevice ::} \DataTypeTok{Device}
\NormalTok{testRemapDevice }\OtherTok{=}\NormalTok{ runMap }\OperatorTok{$}\NormalTok{ remapDevice exampleDevice}
\end{Highlighting}
\end{Shaded}

In other words, in general, one can view
\VERB|\DataTypeTok{Applicative}|-like computations as computations for
generalized multi-stack machines with arbitrary data types and/or
functions as ``stacks''.

In practice, though, simple direct transformations in the style of
\VERB|\DataTypeTok{Applicative}| parsers seem to be the most useful to
us.

\hypertarget{scott-encoded-representation}{%
\section{Scott-encoded
representation}\label{scott-encoded-representation}}

\label{sec:scott}

The LISP-encoding used above is not the only generic representation for
data types, in this section we shall explore the use of Scott-encoding.

Before we start, let us note that while it is trivial to simply
Scott-encode all the pair constructors and destructors in the above
transformations to get more complicated terms with exactly equivalent
semantics~\cite{Malakhovski:2018:EME}, it just complicates things
structurally, and we shall not explore that route.

The interesting question is whether it is possible to remake the above
machinery directly for Scott-encoded representations of the subject data
types

\begin{Shaded}
\begin{Highlighting}[]
\OtherTok{unDeviceScott ::} \DataTypeTok{Device} \OtherTok{->}\NormalTok{ (}\DataTypeTok{Bool} \OtherTok{->} \DataTypeTok{Int} \OtherTok{->} \DataTypeTok{Int} \OtherTok{->}\NormalTok{ c) }\OtherTok{->}\NormalTok{ c}
\NormalTok{unDeviceScott (}\DataTypeTok{Device}\NormalTok{ b x y) f }\OtherTok{=}\NormalTok{ f b x y}

\OtherTok{unBenchmarkScott ::} \DataTypeTok{Benchmark}\NormalTok{ a}
                 \OtherTok{->}\NormalTok{ (a }\OtherTok{->} \DataTypeTok{String} \OtherTok{->}\NormalTok{ a }\OtherTok{->} \DataTypeTok{String} \OtherTok{->}\NormalTok{ c) }\OtherTok{->}\NormalTok{ c}
\NormalTok{unBenchmarkScott (}\DataTypeTok{Benchmark}\NormalTok{ a b c d) f }\OtherTok{=}\NormalTok{ f a b c d}
\end{Highlighting}
\end{Shaded}

\noindent without reaching for anything else. In other words, would not
it be nice if we could work with a Scott-encoded data type
\VERB|\NormalTok{(a }\OtherTok{->}\NormalTok{ b }\OtherTok{->}\NormalTok{ c }\OtherTok{->} \OperatorTok{...} \OtherTok{->}\NormalTok{ z) }\OtherTok{->}\NormalTok{ z}|
as if it was a heterogeneous list of typed values like LISP-encoding is?

Let us start by noticing that we can, in fact, prepend values to
Scott-encoded representations as if they were heterogeneous lists or
tuples

\begin{Shaded}
\begin{Highlighting}[]
\OtherTok{consS ::}\NormalTok{ s}
      \OtherTok{->}\NormalTok{ (a }\OtherTok{->}\NormalTok{ b)}
      \OtherTok{->}\NormalTok{ ((s }\OtherTok{->}\NormalTok{ a) }\OtherTok{->}\NormalTok{ b)}
\NormalTok{consS s ab sa }\OtherTok{=}\NormalTok{ ab (sa s)}
\end{Highlighting}
\end{Shaded}

To see why this prepends \VERB|\NormalTok{s}| to a Scott-encoded
\VERB|\NormalTok{a }\OtherTok{->}\NormalTok{ b}| substitute, for
instance,
\VERB|\NormalTok{x }\OtherTok{->}\NormalTok{ y }\OtherTok{->}\NormalTok{ b}|
for \VERB|\NormalTok{a}|. Note, however, that there are some important
differences. For instance, Scott-encoded data types, unlike LISP-encoded
ones, can not have a generic \VERB|\NormalTok{unconsS}|

\begin{Shaded}
\begin{Highlighting}[]
\OtherTok{unconsS ::}\NormalTok{ ((s }\OtherTok{->}\NormalTok{ a) }\OtherTok{->}\NormalTok{ b) }\OtherTok{->}\NormalTok{ (s, a }\OtherTok{->}\NormalTok{ b)}
\NormalTok{unconsS f }\OtherTok{=}\NormalTok{ (_, _)}
\end{Highlighting}
\end{Shaded}

\noindent as, in general, all the pieces of a Scott-encoded data type
have to be used all at once. This makes most of our previous derivations
unusable. However, very surprisingly, \VERB|\NormalTok{consS}| seems to
be enough.

By prepending \VERB|\NormalTok{s}| to the Scott-encoded data type we can
emulate pretty-printing code above as follows.\footnote{We tried our
  best to make this comprehensible by making the types speak for
  themselves but, arguably, this and the following listings can only be
  really understood by playing with the Literate Haskell version in
  \texttt{ghci}.}

\begin{Shaded}
\begin{Highlighting}[]
\OtherTok{chopS ::}\NormalTok{ ((s }\OtherTok{->}\NormalTok{ a }\OtherTok{->}\NormalTok{ b) }\OtherTok{->}\NormalTok{ c)}
      \OtherTok{->}\NormalTok{ (s }\OtherTok{->}\NormalTok{ a }\OtherTok{->}\NormalTok{ t)}
      \OtherTok{->}\NormalTok{ ((t }\OtherTok{->}\NormalTok{ b) }\OtherTok{->}\NormalTok{ c)}
\NormalTok{chopS i f o }\OtherTok{=}\NormalTok{ i }\OperatorTok{$}\NormalTok{ \textbackslash{}s a }\OtherTok{->}\NormalTok{ o (f s a)}

\NormalTok{depureShowS f r }\OtherTok{=}\NormalTok{ consS [] (f r)}

\OtherTok{showaS ::}\NormalTok{ (r }\OtherTok{->}\NormalTok{ ([}\DataTypeTok{String}\NormalTok{] }\OtherTok{->}\NormalTok{ a }\OtherTok{->}\NormalTok{ b) }\OtherTok{->}\NormalTok{ c)}
       \OtherTok{->}\NormalTok{ (a }\OtherTok{->} \DataTypeTok{String}\NormalTok{)}
       \OtherTok{->}\NormalTok{ (r }\OtherTok{->}\NormalTok{ ([}\DataTypeTok{String}\NormalTok{] }\OtherTok{->}\NormalTok{ b) }\OtherTok{->}\NormalTok{ c)}
\NormalTok{showaS }\OtherTok{=}\NormalTok{ homWrap }\OperatorTok{$}\NormalTok{ \textbackslash{}st f }\OtherTok{->}\NormalTok{ chopS st }\OperatorTok{$}\NormalTok{ \textbackslash{}s a }\OtherTok{->}\NormalTok{ (f a)}\OperatorTok{:}\NormalTok{s}

\NormalTok{runShowS }\OtherTok{=} \FunctionTok{concat} \OperatorTok{.}\NormalTok{ intersperse }\StringTok{" "} \OperatorTok{.} \FunctionTok{reverse} \OperatorTok{.}\NormalTok{ (\textbackslash{}f }\OtherTok{->}\NormalTok{ f }\FunctionTok{id}\NormalTok{)}

\NormalTok{showDeviceS }\OtherTok{=}\NormalTok{ depureShowS unDeviceScott}
  \OtherTok{`showaS`} \FunctionTok{show}
  \OtherTok{`showaS`} \FunctionTok{show}
  \OtherTok{`showaS`} \FunctionTok{show}

\NormalTok{testShowDeviceS }\OtherTok{=}\NormalTok{ runShowS }\OperatorTok{$}\NormalTok{ showDeviceS exampleDevice}
\end{Highlighting}
\end{Shaded}

The only new parts here are the implementation of
\VERB|\NormalTok{chopS}| function, the use of \VERB|\NormalTok{consS}|
instead of the pair constructor, and the replacement of
\VERB|\FunctionTok{fst}| with
\VERB|\NormalTok{\textbackslash{}f }\OtherTok{->}\NormalTok{ f }\FunctionTok{id}|.
The rest is produced mechanically by adding \texttt{S} suffix to all
function calls. The \VERB|\FunctionTok{map}| example can be similarly
mechanically translated as follows.

\begin{Shaded}
\begin{Highlighting}[]
\OtherTok{mapaS ::}\NormalTok{ (r }\OtherTok{->}\NormalTok{ ((x }\OtherTok{->}\NormalTok{ y) }\OtherTok{->}\NormalTok{ a }\OtherTok{->}\NormalTok{ b) }\OtherTok{->}\NormalTok{ c)}
      \OtherTok{->}\NormalTok{ (a }\OtherTok{->}\NormalTok{ x)}
      \OtherTok{->}\NormalTok{ (r }\OtherTok{->}\NormalTok{ (y }\OtherTok{->}\NormalTok{ b) }\OtherTok{->}\NormalTok{ c)}
\NormalTok{mapaS }\OtherTok{=}\NormalTok{ homWrap }\OperatorTok{$}\NormalTok{ \textbackslash{}st f }\OtherTok{->}\NormalTok{ chopS st }\OperatorTok{$}\NormalTok{ \textbackslash{}s a }\OtherTok{->}\NormalTok{ s (f a)}

\NormalTok{depureMapS c f r }\OtherTok{=}\NormalTok{ consS c (f r)}

\NormalTok{runMapS f }\OtherTok{=}\NormalTok{ f }\FunctionTok{id}

\NormalTok{mapDeviceS }\OtherTok{=}\NormalTok{ depureMapS }\DataTypeTok{Device}\NormalTok{ unDeviceScott}
  \OtherTok{`mapaS`} \FunctionTok{not}
  \OtherTok{`mapaS`}\NormalTok{ (}\OperatorTok{+} \DecValTok{100}\NormalTok{)}
  \OtherTok{`mapaS`}\NormalTok{ (}\OperatorTok{+} \DecValTok{200}\NormalTok{)}

\OtherTok{testMapDeviceS ::} \DataTypeTok{Device}
\NormalTok{testMapDeviceS }\OtherTok{=}\NormalTok{ runMapS }\OperatorTok{$}\NormalTok{ mapDeviceS exampleDevice}
\end{Highlighting}
\end{Shaded}

The most interesting part, however, is the reimplementation of
\VERB|\FunctionTok{zip}|. By following the terms in the previous section
we would arrive at the following translation for
\VERB|\NormalTok{depureZip}|

\begin{Shaded}
\begin{Highlighting}[]
\OtherTok{depureZipS' ::}\NormalTok{ s }\OtherTok{->}\NormalTok{ (ra }\OtherTok{->}\NormalTok{ a) }\OtherTok{->}\NormalTok{ (rb }\OtherTok{->}\NormalTok{ b }\OtherTok{->}\NormalTok{ c)}
            \OtherTok{->}\NormalTok{ ra }\OtherTok{->}\NormalTok{ rb}
            \OtherTok{->}\NormalTok{ (s }\OtherTok{->}\NormalTok{ a }\OtherTok{->}\NormalTok{ b) }\OtherTok{->}\NormalTok{ c}
\NormalTok{depureZipS' c f g r s }\OtherTok{=}\NormalTok{ consS c (consS (f r) (g s))}
\end{Highlighting}
\end{Shaded}

\noindent Frustratingly, there is no \VERB|\NormalTok{chop2}| equivalent
for it

\begin{Shaded}
\begin{Highlighting}[]
\OtherTok{chop2S' ::}\NormalTok{ ((s }\OtherTok{->}\NormalTok{ ((a }\OtherTok{->}\NormalTok{ b) }\OtherTok{->}\NormalTok{ c) }\OtherTok{->}\NormalTok{ d }\OtherTok{->}\NormalTok{ e) }\OtherTok{->}\NormalTok{ f)}
        \OtherTok{->}\NormalTok{ (s }\OtherTok{->}\NormalTok{ a }\OtherTok{->}\NormalTok{ d }\OtherTok{->}\NormalTok{ t)}
        \OtherTok{->}\NormalTok{ (t }\OtherTok{->}\NormalTok{ (b }\OtherTok{->}\NormalTok{ c) }\OtherTok{->}\NormalTok{ e) }\OtherTok{->}\NormalTok{ f}
\NormalTok{chop2S' i f o }\OtherTok{=}\NormalTok{ i }\OperatorTok{$}\NormalTok{ \textbackslash{}s abq d }\OtherTok{->}\NormalTok{ o _ _}
\end{Highlighting}
\end{Shaded}

\noindent because \VERB|\NormalTok{a}| becomes effectively inaccessible
in this order of \VERB|\NormalTok{consS}|ing (as there is no
\VERB|\NormalTok{unconsS}|). However, fascinatingly, by simply changing
that order to

\begin{Shaded}
\begin{Highlighting}[]
\NormalTok{depureZipS c f g r s }\OtherTok{=}\NormalTok{ consS (consS c (f r)) (g s)}
\end{Highlighting}
\end{Shaded}

\noindent we get our \VERB|\NormalTok{cons2S}| and, by mechanical
translation, all the rest of \VERB|\NormalTok{zipDevice}| example

\begin{Shaded}
\begin{Highlighting}[]
\OtherTok{chop2S ::}\NormalTok{ ((((s }\OtherTok{->}\NormalTok{ a }\OtherTok{->}\NormalTok{ b) }\OtherTok{->}\NormalTok{ c) }\OtherTok{->}\NormalTok{ d }\OtherTok{->}\NormalTok{ e) }\OtherTok{->}\NormalTok{ f)}
       \OtherTok{->}\NormalTok{ (s }\OtherTok{->}\NormalTok{ a }\OtherTok{->}\NormalTok{ d }\OtherTok{->}\NormalTok{ t)}
       \OtherTok{->}\NormalTok{ (((t }\OtherTok{->}\NormalTok{ b) }\OtherTok{->}\NormalTok{ c) }\OtherTok{->}\NormalTok{ e) }\OtherTok{->}\NormalTok{ f}
\NormalTok{chop2S i f o }\OtherTok{=}\NormalTok{ i }\OperatorTok{$}\NormalTok{ \textbackslash{}sabc d }\OtherTok{->}\NormalTok{ o }\OperatorTok{$}\NormalTok{ \textbackslash{}tb }\OtherTok{->}\NormalTok{ sabc }\OperatorTok{$}\NormalTok{ \textbackslash{}s a }\OtherTok{->}\NormalTok{ tb }\OperatorTok{$}\NormalTok{ f s a d}

\OtherTok{zipaS ::}\NormalTok{ (ra }\OtherTok{->}\NormalTok{ rb }\OtherTok{->}\NormalTok{ (((((x }\OtherTok{->}\NormalTok{ y) }\OtherTok{->}\NormalTok{ a }\OtherTok{->}\NormalTok{ b) }\OtherTok{->}\NormalTok{ c) }\OtherTok{->}\NormalTok{ d }\OtherTok{->}\NormalTok{ e) }\OtherTok{->}\NormalTok{ f))}
      \OtherTok{->}\NormalTok{ (a }\OtherTok{->}\NormalTok{ d }\OtherTok{->}\NormalTok{ x)}
      \OtherTok{->}\NormalTok{ (ra }\OtherTok{->}\NormalTok{ rb }\OtherTok{->}\NormalTok{ (((y }\OtherTok{->}\NormalTok{ b) }\OtherTok{->}\NormalTok{ c) }\OtherTok{->}\NormalTok{ e) }\OtherTok{->}\NormalTok{ f)}
\NormalTok{zipaS }\OtherTok{=}\NormalTok{ homWrap2 }\OperatorTok{$}\NormalTok{ \textbackslash{}st f }\OtherTok{->}\NormalTok{ chop2S st }\OperatorTok{$}\NormalTok{ \textbackslash{}s a b }\OtherTok{->}\NormalTok{ s (f a b)}

\NormalTok{runZipS f }\OtherTok{=}\NormalTok{ f }\FunctionTok{id} \FunctionTok{id}

\NormalTok{zipDeviceS }\OtherTok{=}\NormalTok{ depureZipS }\DataTypeTok{Device}\NormalTok{ unDeviceScott unDeviceScott}
  \OtherTok{`zipaS`}\NormalTok{ (}\OperatorTok{&&}\NormalTok{)}
  \OtherTok{`zipaS`}\NormalTok{ (}\OperatorTok{+}\NormalTok{)}
  \OtherTok{`zipaS`}\NormalTok{ (}\OperatorTok{+}\NormalTok{)}

\OtherTok{testZipDeviceS ::} \DataTypeTok{Device}
\NormalTok{testZipDeviceS }\OtherTok{=}\NormalTok{ runZipS }\OperatorTok{$}\NormalTok{ zipDeviceS exampleDevice testMapDeviceS}
\end{Highlighting}
\end{Shaded}

\noindent thus, again, implementing all the examples from
\cref{sec:examples}, but now purely with Scott-encoded data types.

\begin{remark}

Note that while the transformation form \VERB|\NormalTok{b}| to
\VERB|\NormalTok{(a, b)}| for the LISP-encoding or the plain tuples is
regular, the transformation from
\VERB|\NormalTok{(a }\OtherTok{->}\NormalTok{ b }\OtherTok{->}\NormalTok{ c }\OtherTok{->} \OperatorTok{...} \OtherTok{->}\NormalTok{ z) }\OtherTok{->}\NormalTok{ z}|
to
\VERB|\NormalTok{(s }\OtherTok{->}\NormalTok{ a }\OtherTok{->}\NormalTok{ b }\OtherTok{->}\NormalTok{ c }\OtherTok{->} \OperatorTok{...} \OtherTok{->}\NormalTok{ z) }\OtherTok{->}\NormalTok{ z}|
is not, the former is not a sub-expression of the latter. Taking that
into account, we feel that the very fact that the implementations
demonstrated above are even possible is rather fascinating. The fact
that Scott-encoding can be used as a heterogeneous list is rather
surprising as even the fact that \VERB|\NormalTok{consS}| is possible is
rather weird, not to mention the fact that useful things can be done
without \VERB|\NormalTok{unconsS}|. We are not aware of any literature
that describes doing anything similar directly to Scott-encoded data
types.

\end{remark}

\hypertarget{general-case}{%
\section{General Case}\label{general-case}}

\label{sec:general-case}

Curiously, note that with the aforementioned order of
\VERB|\NormalTok{consS}|ing \VERB|\NormalTok{chop2S}| is actually a
special case of \VERB|\NormalTok{chopS}|

\begin{Shaded}
\begin{Highlighting}[]
\OtherTok{chop2S' ::}\NormalTok{ ((((s }\OtherTok{->}\NormalTok{ a }\OtherTok{->}\NormalTok{ b) }\OtherTok{->}\NormalTok{ c) }\OtherTok{->}\NormalTok{ d }\OtherTok{->}\NormalTok{ e) }\OtherTok{->}\NormalTok{ f)}
        \OtherTok{->}\NormalTok{ (s }\OtherTok{->}\NormalTok{ a }\OtherTok{->}\NormalTok{ d }\OtherTok{->}\NormalTok{ t)}
        \OtherTok{->}\NormalTok{ (((t }\OtherTok{->}\NormalTok{ b) }\OtherTok{->}\NormalTok{ c) }\OtherTok{->}\NormalTok{ e) }\OtherTok{->}\NormalTok{ f}
\NormalTok{chop2S' i f o }\OtherTok{=}\NormalTok{ chopS i (\textbackslash{}sabc d tb }\OtherTok{->}\NormalTok{ sabc }\OperatorTok{$}\NormalTok{ \textbackslash{}s a }\OtherTok{->}\NormalTok{ tb }\OperatorTok{$}\NormalTok{ f s a d) o}
  \CommentTok{-- == chop2S}
\end{Highlighting}
\end{Shaded}

\noindent and this pattern continues when \VERB|\NormalTok{consS}|ing
more structures

\begin{Shaded}
\begin{Highlighting}[]
\OtherTok{depureZip3S ::}\NormalTok{ s }\OtherTok{->}\NormalTok{ (ra }\OtherTok{->}\NormalTok{ a }\OtherTok{->}\NormalTok{ b) }\OtherTok{->}\NormalTok{ (rb }\OtherTok{->}\NormalTok{ c }\OtherTok{->}\NormalTok{ d) }\OtherTok{->}\NormalTok{ (rc }\OtherTok{->}\NormalTok{ e }\OtherTok{->}\NormalTok{ f)}
            \OtherTok{->}\NormalTok{ ra }\OtherTok{->}\NormalTok{ rb }\OtherTok{->}\NormalTok{ rc}
            \OtherTok{->}\NormalTok{ (((((s }\OtherTok{->}\NormalTok{ a) }\OtherTok{->}\NormalTok{ b) }\OtherTok{->}\NormalTok{ c) }\OtherTok{->}\NormalTok{ d) }\OtherTok{->}\NormalTok{ e) }\OtherTok{->}\NormalTok{ f}
\NormalTok{depureZip3S c f g h r s t }\OtherTok{=}\NormalTok{ consS (consS (consS c (f r)) (g s)) (h t)}

\OtherTok{chop3S ::}\NormalTok{ ((((((s }\OtherTok{->}\NormalTok{ a }\OtherTok{->}\NormalTok{ b) }\OtherTok{->}\NormalTok{ c) }\OtherTok{->}\NormalTok{ d }\OtherTok{->}\NormalTok{ e) }\OtherTok{->}\NormalTok{ f) }\OtherTok{->}\NormalTok{ g }\OtherTok{->}\NormalTok{ h) }\OtherTok{->}\NormalTok{ i)}
       \OtherTok{->}\NormalTok{ (s }\OtherTok{->}\NormalTok{ a }\OtherTok{->}\NormalTok{ d }\OtherTok{->}\NormalTok{ g }\OtherTok{->}\NormalTok{ t)}
       \OtherTok{->}\NormalTok{ (((((t }\OtherTok{->}\NormalTok{ b) }\OtherTok{->}\NormalTok{ c) }\OtherTok{->}\NormalTok{ e) }\OtherTok{->}\NormalTok{ f) }\OtherTok{->}\NormalTok{ h) }\OtherTok{->}\NormalTok{ i}
\NormalTok{chop3S i f o }\OtherTok{=}\NormalTok{ chop2S i (\textbackslash{}sabc d g tb }\OtherTok{->}\NormalTok{ sabc }\OperatorTok{$}\NormalTok{ \textbackslash{}s a }\OtherTok{->}\NormalTok{ tb }\OperatorTok{$}\NormalTok{ f s a d g) o}

\CommentTok{-- and so on}
\end{Highlighting}
\end{Shaded}

The same is true for LISP-encoded variant since we can use the same
order of \VERB|\NormalTok{cons}|ing there, e.g.

\begin{Shaded}
\begin{Highlighting}[]
\OtherTok{chop2' ::}\NormalTok{ ((s, (a, b)), (c, d))}
       \OtherTok{->}\NormalTok{ (s }\OtherTok{->}\NormalTok{ a }\OtherTok{->}\NormalTok{ c }\OtherTok{->}\NormalTok{ t)}
       \OtherTok{->}\NormalTok{ ((t, b), d)}
\NormalTok{chop2' (sab, (c, d)) f }\OtherTok{=}\NormalTok{ (chop sab (\textbackslash{}s a }\OtherTok{->}\NormalTok{ f s a c), d)}
  \CommentTok{-- = chop2}
\end{Highlighting}
\end{Shaded}

\noindent but we think this presentation makes things look more complex
there, not less. Though, as we shall see in the next section (in its
Literal Haskell version), we could have simplified the general case by
using \VERB|\NormalTok{chop2'}| above.

In other words, if we are to \VERB|\NormalTok{cons}| LISP-encoded and
\VERB|\NormalTok{consS}| Scott-encoded data types in the right order
then all of the \VERB|\DataTypeTok{Applicative}|-like operators of this
paper and the generalizations of \VERB|\DataTypeTok{Applicative}|-like
\VERB|\FunctionTok{zip}|s to larger numbers of structures can be
uniformly produced from just \VERB|\NormalTok{chop}| and
\VERB|\NormalTok{chopS}|.

\hypertarget{formal-account}{%
\section{Formal Account}\label{formal-account}}

\label{sec:formally}

The derivation of \cref{sec:deriving-the-technique}, as demonstrated by
the following sections, describes a technique (as opposed to an isolated
example) for expressing transformations between simple data types of a
single constructor using \VERB|\DataTypeTok{Applicative}|-like
computations. More formally, that technique consists of

\begin{itemize}
\item
  deconstructing the data type (into its LISP-encoded representation in
  \cref{sec:deriving-the-technique,sec:implementation} or Scott-encoded
  representation in \cref{sec:scott}),
\item
  wrapping the deconstructed representation into the
  \VERB|\DataTypeTok{Applicative}|-like structure in question with an
  operation analogous to \VERB|\DataTypeTok{Applicative}|'s
  \VERB|\FunctionTok{pure}| (\VERB|\NormalTok{depureShow}|, etc),
\item
  followed by spelling out transformation steps to the desired
  representation by interspersing them with an operator analogous to
  \VERB|\DataTypeTok{Applicative}|'s
  \VERB|\NormalTok{(}\OperatorTok{<*>}\NormalTok{)}|
  (\VERB|\NormalTok{showa}|, \VERB|\NormalTok{mapa}|,
  \VERB|\NormalTok{zipa}|, etc),
\item
  followed by wrapping the whole structure into
  \VERB|\NormalTok{(}\OtherTok{->}\NormalTok{) r}|
  \VERB|\DataTypeTok{Reader}| that is used to propagate the input
  argument to the front of the expression without adding explicit
  argument bindings to the whole expressions.
\end{itemize}

Note, however, that the last ``wrapping'' bit of the translation is
orthogonal to the rest. It is needed to produce a completely
variable-binding-less expression, but that step can be skipped if
variable-binding-lessness is not desired: one simply needs to remove the
\VERB|\NormalTok{homWrap}| wrapping, add an explicitly bound argument to
the function, and then apply it to \VERB|\NormalTok{depureShow}|.

Also remember that \cref{sec:implementation} showed that, in general,
those expressions can implement any computations for generalized
multi-stack machines with arbitrary data types and/or functions as
``stacks''. For the \VERB|\FunctionTok{show}|-,
\VERB|\FunctionTok{map}|-, and \VERB|\FunctionTok{zip}|-like
transformations we described in detail, however, the central
\VERB|\NormalTok{chop}| operator corresponds to a simple state
transformer of the corresponding ``step-by-step''
\VERB|\NormalTok{fold}|, if we are to view the deconstructed data type
as a heterogeneous list.

Finally, note that while \VERB|\NormalTok{depureMap}| and
\VERB|\NormalTok{depureZip}| (\VERB|\NormalTok{depureMapS}| and
\VERB|\NormalTok{depureZipS}|) take more arguments than
\VERB|\DataTypeTok{Applicative}|'s \VERB|\FunctionTok{pure}| this fact
is actually inconsequential as in \cref{sec:general-case} we noted that
we can simply reorganize all our expressions to \VERB|\NormalTok{cons}|
to the left (as we had to do for Scott-encoded data types). Thus, only
the last argument to the \VERB|\NormalTok{depure}\OperatorTok{*}|
functions is of any consequence to the general structure (since it is
the argument we are \VERB|\NormalTok{fold}|ing on, inductively
speaking), the rest are simply baggage used internally by the
corresponding operators.

\hypertarget{dependently-typed-applicative}{%
\subsection{Dependently-typed
Applicative}\label{dependently-typed-applicative}}

Now, the obvious question is how a general structure unifying all those
operators would look. Firstly, let us note that the
\VERB|\FunctionTok{pure}| function of \VERB|\DataTypeTok{Applicative}|
can be separated out into its own type class

\begin{Shaded}
\begin{Highlighting}[]
\KeywordTok{class} \DataTypeTok{Pointed}\NormalTok{ f }\KeywordTok{where}
\OtherTok{  pure ::}\NormalTok{ a }\OtherTok{->}\NormalTok{ f a}

\KeywordTok{infixl} \DecValTok{4} \OperatorTok{<*>}
\KeywordTok{class}\NormalTok{ (}\DataTypeTok{Pointed}\NormalTok{ f, }\DataTypeTok{Functor}\NormalTok{ f) }\OtherTok{=>} \DataTypeTok{Applicative}\NormalTok{ f }\KeywordTok{where}
\OtherTok{  (<*>) ::}\NormalTok{ f (a }\OtherTok{->}\NormalTok{ b) }\OtherTok{->}\NormalTok{ f a }\OtherTok{->}\NormalTok{ f b}
\end{Highlighting}
\end{Shaded}

\noindent Moreover, note that, algebraically speaking,
\VERB|\DataTypeTok{Applicative}| depends on \VERB|\DataTypeTok{Pointed}|
only because their combination gives \VERB|\DataTypeTok{Functor}|, they
are independent otherwise. Since we have no equivalent for
\VERB|\DataTypeTok{Functor}| with \VERB|\DataTypeTok{Applicative}|-like
expressions we can discuss these two parts separately.

Secondly, let us note that \VERB|\DataTypeTok{Control.Category}| and
\VERB|\DataTypeTok{Control.Arrow}| modules of
\texttt{base}~\cite{Hackage:base4900} define
\VERB|\DataTypeTok{Category}|~\cite{Hackage:base4900} and
\VERB|\DataTypeTok{ArrowApply}|~\cite{hughes-arrows-00} type classes as

\begin{Shaded}
\begin{Highlighting}[]
\KeywordTok{class} \DataTypeTok{Category}\NormalTok{ cat }\KeywordTok{where}
\OtherTok{  id ::}\NormalTok{ cat a a}
\OtherTok{  (.) ::}\NormalTok{ cat b c }\OtherTok{->}\NormalTok{ cat a b }\OtherTok{->}\NormalTok{ cat a c}

\KeywordTok{class} \DataTypeTok{Arrow}\NormalTok{ a }\OtherTok{=>} \DataTypeTok{ArrowApply}\NormalTok{ a }\KeywordTok{where}
\OtherTok{  app ::}\NormalTok{ a (a b c, b) c}
\end{Highlighting}
\end{Shaded}

\noindent respectively. Both of these type classes denote generalized
functions over generalized function types: \VERB|\NormalTok{cat}| and
\VERB|\NormalTok{a}| respectively.

Thirdly, if we are to look at the types of our \VERB|\NormalTok{showa}|,
\VERB|\NormalTok{mapa}|, and \VERB|\NormalTok{zipa}| operators and their
versions for Scott-encoded data types, the most glaring difference from
the type of \VERB|\NormalTok{(}\OperatorTok{<*>}\NormalTok{)}| we will
notice is the fact that the types of their second arguments and the
types of their results depend on the types of their first arguments (or,
equivalently, we can say that all of those depend on another implicit
type argument). In other words, if
\VERB|\NormalTok{(}\OperatorTok{<*>}\NormalTok{)}| and
\VERB|\NormalTok{app}| are two generalizations of the conventional
function application, then the structure that describes our operators is
a generalization of the dependently typed function application.

The simplest general encoding we have for our examples for GHC Haskell
(with awful lot of extensions) looks like this

\begin{Shaded}
\begin{Highlighting}[]
\KeywordTok{class} \DataTypeTok{ApplicativeLike}\NormalTok{ f }\KeywordTok{where}
  \KeywordTok{type} \DataTypeTok{C}\NormalTok{ f a}\OtherTok{ b ::} \OperatorTok{*} \CommentTok{-- type of arrow under `f`}
  \KeywordTok{type} \DataTypeTok{G}\NormalTok{ f}\OtherTok{ a ::} \OperatorTok{*}   \CommentTok{-- type of argument dependent on `f`}
  \KeywordTok{type} \DataTypeTok{F}\NormalTok{ f}\OtherTok{ b ::} \OperatorTok{*}   \CommentTok{-- type of result dependent on `f`}
\OtherTok{  (<**>) ::}\NormalTok{ f (}\DataTypeTok{C}\NormalTok{ f a b) }\OtherTok{->} \DataTypeTok{G}\NormalTok{ f a }\OtherTok{->} \DataTypeTok{F}\NormalTok{ f b}

\KeywordTok{newtype} \DataTypeTok{Mapper}\NormalTok{ r f a }\OtherTok{=} \DataTypeTok{Mapper}\NormalTok{ \{}\OtherTok{ runMapper ::}\NormalTok{ r }\OtherTok{->}\NormalTok{ (f, a) \}}

\KeywordTok{instance} \DataTypeTok{ApplicativeLike}\NormalTok{ (}\DataTypeTok{Mapper}\NormalTok{ e (x }\OtherTok{->}\NormalTok{ y)) }\KeywordTok{where}
  \KeywordTok{type} \DataTypeTok{C}\NormalTok{ (}\DataTypeTok{Mapper}\NormalTok{ e (x }\OtherTok{->}\NormalTok{ y)) a b }\OtherTok{=}\NormalTok{ (a, b)}
  \KeywordTok{type} \DataTypeTok{G}\NormalTok{ (}\DataTypeTok{Mapper}\NormalTok{ e (x }\OtherTok{->}\NormalTok{ y)) a }\OtherTok{=}\NormalTok{ a }\OtherTok{->}\NormalTok{ x}
  \KeywordTok{type} \DataTypeTok{F}\NormalTok{ (}\DataTypeTok{Mapper}\NormalTok{ e (x }\OtherTok{->}\NormalTok{ y)) b }\OtherTok{=} \DataTypeTok{Mapper}\NormalTok{ e y b}
\NormalTok{  f }\OperatorTok{<**>}\NormalTok{ g }\OtherTok{=} \DataTypeTok{Mapper} \OperatorTok{$}\NormalTok{ mapa (runMapper f) g}

\OtherTok{mapDeviceG ::} \DataTypeTok{Mapper} \DataTypeTok{Device} \DataTypeTok{Device}\NormalTok{ ()}
\NormalTok{mapDeviceG }\OtherTok{=} \DataTypeTok{Mapper}\NormalTok{ (depureMap }\DataTypeTok{Device}\NormalTok{ unDeviceLISP)}
  \OperatorTok{<**>} \FunctionTok{not}
  \OperatorTok{<**>}\NormalTok{ (}\OperatorTok{+} \DecValTok{100}\NormalTok{)}
  \OperatorTok{<**>}\NormalTok{ (}\OperatorTok{+} \DecValTok{200}\NormalTok{)}

\OtherTok{testMapDeviceG ::} \DataTypeTok{Device}
\NormalTok{testMapDeviceG }\OtherTok{=}\NormalTok{ runMap }\OperatorTok{$}\NormalTok{ runMapper mapDeviceG exampleDevice}

\KeywordTok{newtype} \DataTypeTok{MapperS}\NormalTok{ c r f a }\OtherTok{=} \DataTypeTok{MapperS}
\NormalTok{  \{}\OtherTok{ runMapperS ::}\NormalTok{ r }\OtherTok{->}\NormalTok{ (f }\OtherTok{->}\NormalTok{ a) }\OtherTok{->}\NormalTok{ c \}}

\KeywordTok{instance} \DataTypeTok{ApplicativeLike}\NormalTok{ (}\DataTypeTok{MapperS}\NormalTok{ c e (x }\OtherTok{->}\NormalTok{ y)) }\KeywordTok{where}
  \KeywordTok{type} \DataTypeTok{C}\NormalTok{ (}\DataTypeTok{MapperS}\NormalTok{ c e (x }\OtherTok{->}\NormalTok{ y)) a b }\OtherTok{=}\NormalTok{ a }\OtherTok{->}\NormalTok{ b}
  \KeywordTok{type} \DataTypeTok{G}\NormalTok{ (}\DataTypeTok{MapperS}\NormalTok{ c e (x }\OtherTok{->}\NormalTok{ y)) a }\OtherTok{=}\NormalTok{ a }\OtherTok{->}\NormalTok{ x}
  \KeywordTok{type} \DataTypeTok{F}\NormalTok{ (}\DataTypeTok{MapperS}\NormalTok{ c e (x }\OtherTok{->}\NormalTok{ y)) b }\OtherTok{=} \DataTypeTok{MapperS}\NormalTok{ c e y b}
\NormalTok{  f }\OperatorTok{<**>}\NormalTok{ g }\OtherTok{=} \DataTypeTok{MapperS} \OperatorTok{$}\NormalTok{ mapaS (runMapperS f) g}

\OtherTok{mapDeviceGS ::} \DataTypeTok{MapperS}\NormalTok{ c }\DataTypeTok{Device} \DataTypeTok{Device}\NormalTok{ c}
\NormalTok{mapDeviceGS }\OtherTok{=} \DataTypeTok{MapperS}\NormalTok{ (depureMapS }\DataTypeTok{Device}\NormalTok{ unDeviceScott)}
  \OperatorTok{<**>} \FunctionTok{not}
  \OperatorTok{<**>}\NormalTok{ (}\OperatorTok{+} \DecValTok{100}\NormalTok{)}
  \OperatorTok{<**>}\NormalTok{ (}\OperatorTok{+} \DecValTok{200}\NormalTok{)}

\OtherTok{testMapDeviceGS ::} \DataTypeTok{Device}
\NormalTok{testMapDeviceGS }\OtherTok{=}\NormalTok{ runMapS }\OperatorTok{$}\NormalTok{ runMapperS mapDeviceGS exampleDevice}

\CommentTok{-- See Literate Haskell version for many more examples.}
\end{Highlighting}
\end{Shaded}

The operator analogous to \VERB|\FunctionTok{pure}| simply wraps the
result produced by the data type destructor into the corresponding
initial state, thus its generalization is not interesting (in general,
it is a function \VERB|\NormalTok{a }\OtherTok{->}\NormalTok{ f b}|).
Moreover, generalizing it actually adds problems because a generic
\VERB|\NormalTok{depure}| makes
\VERB|\NormalTok{(}\OperatorTok{<**>}\NormalTok{)}| ambitious in

\begin{Shaded}
\begin{Highlighting}[]
\NormalTok{ambitiousExample a }\OtherTok{=}\NormalTok{ depure unDevice }\OperatorTok{<**>}\NormalTok{ a }\OperatorTok{<**>}\NormalTok{ a }\OperatorTok{<**>}\NormalTok{ a}
\end{Highlighting}
\end{Shaded}

\noindent This does not happen for \VERB|\DataTypeTok{Applicative}| type
class since both arguments to
\VERB|\NormalTok{(}\OperatorTok{<*>}\NormalTok{)}| are of the same type
family \VERB|\NormalTok{f}| there.

\hypertarget{conclusion}{%
\section{Conclusion}\label{conclusion}}

\label{sec:conclusion}

From a practical perspective, in this article we have shown that by
implementing a series of rather trivial state transformers we called
\VERB|\NormalTok{chop}\OperatorTok{*}| and wrappers into a
\VERB|\NormalTok{(}\OtherTok{->}\NormalTok{) r}|
\VERB|\DataTypeTok{Reader}| we called
\VERB|\NormalTok{homWrap}\OperatorTok{*}| and then composing them one
can express operators that can implement arbitrary computations for
generalized multi-stack machines using a rather curious form of
expressions very similar to conventional
\VERB|\DataTypeTok{Applicative}| parsers. Then, we demonstrated how to
use those operators to implement \VERB|\DataTypeTok{Applicative}|-like
pretty-printers, \VERB|\FunctionTok{map}|s, and
\VERB|\FunctionTok{zip}|s between simple data types of a single
constructor by first unfolding them into LISP- and Scott-encoded
representations and then folding them back with custom ``step-by-step''
\VERB|\NormalTok{fold}|s. (Where the very fact that Scott-encoded case
is even possible is rather fascinating as those terms are constructed
using a rather unorthodox technique.)

\begin{remark}

By the way, note that Haskell's
\VERB|\DataTypeTok{GHC.Generics}|~\cite{GHC86:base412:Generics} is not
an adequate replacement for LISP- and Scott-encoded representations used
in the paper: not only is the \VERB|\DataTypeTok{Rep}| type family
complex, its structure is not even deterministic as GHC tries to keep
the resulting type representation tree balanced. Which, practically
speaking, suggests another \VERB|\DataTypeTok{GHC}| extension.

\end{remark}

From a theoretical perspective, in this article we have presented a
natural generalization of the conventional
\VERB|\DataTypeTok{Applicative}|\cite{mcbride-paterson-08} type class
(which can be viewed as a generalization of conventional function
application) into dependent types with generalized arrow of
\VERB|\DataTypeTok{Category}|/\VERB|\DataTypeTok{ArrowApply}|~\cite{Hackage:base4900, hughes-arrows-00}.
Both \VERB|\DataTypeTok{Applicative}|s and
\VERB|\DataTypeTok{Monad}|s~\cite{moggi-89, moggi-91, Wadler:1992:EFP}
(that can be viewed as a generalization of the conventional sequential
composition of actions, aka ``imperative semicolon'') were similarly
generalized to superapplicatives and supermonads
in~\cite{Bracker2018:SS}. In particular, \cite{Bracker2018:SS} starts by
giving the following definition for \VERB|\DataTypeTok{Applicative}|

\begin{Shaded}
\begin{Highlighting}[]
\KeywordTok{class} \DataTypeTok{Applicative}\NormalTok{ m n p }\KeywordTok{where}
\OtherTok{  (<*>) ::}\NormalTok{ m (a }\OtherTok{->}\NormalTok{ b) }\OtherTok{->}\NormalTok{ n a }\OtherTok{->}\NormalTok{ p b}
\end{Highlighting}
\end{Shaded}

\noindent then adds constraints on top to make the type inference work,
and then requires all of \VERB|\NormalTok{m}|, \VERB|\NormalTok{n}|, and
\VERB|\NormalTok{p}| to be \VERB|\DataTypeTok{Functor}|s (producing such
a long and scary type class signature as the result so that we decided
against including it here). In contrast, our
\VERB|\DataTypeTok{ApplicativeLike}| generalizes the arrow under
\VERB|\NormalTok{m}|, goes straight to dependent types for
\VERB|\NormalTok{n}| and \VERB|\NormalTok{p}| instead of ad-hoc
constraints, and doesn't constrain them in any other way.

\begin{remark}

Which suggests syntactic (rather than algebraic) treatment of
\VERB|\DataTypeTok{ApplicativeLike}| structure as it seems that there
are no new interesting laws about it except for those that are true for
the conventional function application (e.g., congruence
\VERB|\NormalTok{a }\OperatorTok{==}\NormalTok{ b }\OtherTok{=>}\NormalTok{ f a }\OperatorTok{==}\NormalTok{ f b}|).

\end{remark}

In other words, our \VERB|\DataTypeTok{ApplicativeLike}| can be viewed
as a simpler encoding for generalized superapplicatives
of~\cite{Bracker2018:SS} when those are treated syntactically rather
than algebraically (since we completely ignore
\VERB|\DataTypeTok{Functor}|s).

Future fork on the subject consists of applying the same ideas to
\VERB|\DataTypeTok{Alternative}| type class to cover the
multi-constructor case, which is not clear at the moment since it is not
exactly clear how the canonical use of \VERB|\DataTypeTok{Alternative}|
for parsing tagged data types should look like in the first place, as,
unlike the \VERB|\DataTypeTok{Applicative}| case, different libraries
use different idioms for this.

\printbibliography

\end{document}